\documentclass{article}

\usepackage{PRIMEarxiv}
\usepackage{textcomp}

\usepackage{graphicx}
\usepackage{amsmath}
\usepackage[version=4]{mhchem}
\usepackage{siunitx}
\usepackage{longtable,tabularx}
\usepackage{tikz}
\usepackage{pgfplots}
\usepackage{geometry}
\usepackage{soul}
\usepackage{float}

\usepackage[utf8]{inputenc} % allow utf-8 input
\usepackage[T1]{fontenc}    % use 8-bit T1 fonts
\usepackage{hyperref}       % hyperlinks
\usepackage{url}            % simple URL typesetting
\usepackage{booktabs}       % professional-quality tables
\usepackage{amsfonts}       % blackboard math symbols
\usepackage{nicefrac}       % compact symbols for 1/2, etc.
\usepackage{microtype}      % microtypography
\usepackage{lipsum}
\usepackage{fancyhdr}       % header
\graphicspath{{media/}}     % organize your images and other figures under media/ folder

\usepackage{lineno}
\usepackage{amsmath}
\usepackage{amssymb}
\usepackage{algorithm}
\usepackage{algpseudocode}
\usepackage{subcaption}
\usepackage{tikz}
\usepackage{booktabs}
\usepackage{multirow}
\usepackage{tabularx}

\modulolinenumbers[5]

\usetikzlibrary{3d,calc}

\title{Nonlinear Propagation of Non-Gaussian Uncertainties
}

\author{
  Giacomo Acciarini \\
  University of Surrey \\
  \texttt{g.acciarini@surrey.ac.uk} \\
   \And
   Nicola Baresi\\
   University of Surrey\\
   \texttt{nb0039@surrey.ac.uk}\\
   \AND
   David Lloyd\\
   University of Surrey\\
   \texttt{d.lloyd@surrey.ac.uk}\\
  \And
  Dario Izzo \\
  European Space Agency \\
  \texttt{dario.izzo@esa.int}
}
\DeclareMathOperator{\EX}{\mathbb{E}}

\pgfplotsset{compat=1.18}

\begin{document}
\let\oldhat\hat
\renewcommand{\vec}[1]{\pmb{\mathrm{#1}}}
\renewcommand{\hat}[1]{\oldhat{\pmb{\mathrm{#1}}}}
\maketitle
\keywords{nonlinear uncertainty propagation \and non-Gaussian uncertainties \and moment generating functions \and MGF \and astrodynamics \and state transition tensors \and event transition tensors \and high-order variational equations \and differential algebra \and generalized dual numbers \and Fokker-Planck Equation}
\section{Introduction}
Uncertainty propagation plays a crucial role in various aspects of space trajectory design, including low-Earth orbit determination, satellite tracking, collision avoidance, robust trajectory estimation in deep space missions, and space situational awareness~\cite{luo2017review}. Approaches to uncertainty propagation vary widely, from attempts to solve the Fokker-Planck equation directly~\cite{giza2009approach, acciarini2024uncertainty}, using approximated or numerical methods, to techniques based on first-order dynamics expansions (e.g., linear covariance propagation)\cite{butcher2017kalman}, polynomial chaos expansion\cite{jones2013nonlinear}, and change of variable technique~\cite{wittig2017long}. Other methods leverage statistical properties of Keplerian orbits~\cite{dennis1972probabilistic, izzo2005effects} or apply Gaussian mixture models combined with unscented transformations~\cite{vishwajeet2014nonlinear}.

One of the most popular methods is via state transition tensors (STT), which leverages the high-order expansion of the flow around a nominal trajectory to map uncertainties at future times~\cite{park2006nonlinear, fujimoto2012analytical}. However, methods based on STTs have so far only dealt with initial Gaussian distributions, thereby preventing the treatment of generic probability density functions on both initial conditions and problem parameters.

While STTs can be computed via the numerical integration of the variational equations, different approaches have proposed the use of alternative techniques to perform this task. These range from generalized dual numbers~\cite{izzo2018audi} to differential algebra~\cite{berz1998verified, armellin2010asteroid} to machine learning pipelines as popularized, for example, by the recent NeuralODE paradigm~\cite{chen2018neuralode, jax2018github}. According to the context,  STTs are often referred to using a variety of different names and notations.
We will only use the term state transition tensors, referring to the quantities that represent the high-order derivatives of the flow with respect to the initial conditions and problem parameters. 
With reference to the work of Park \& Scheeres~\cite{park2006nonlinear}, we extend the ideas on uncertainty propagation through STTs to the case of non-Gaussian probability density functions (PDFs) associated to the dynamical system initial conditions and parameters (e.g., initial position, gravitational constants, etc.).
 
In the past, nonlinear uncertainty propagation has been primarily studied in the case of distributions that are initially either Gaussian or described by mixtures of Gaussian~\cite{khatri2024hybrid, losacco2024low, boone2023directional, massari2017nonlinear, yang2019nonlinear, sun2019nonlinear}. Our approach allows dropping the Gaussian assumption, as generic mappings for any type of PDF can be found, provided that the PDF admits a moment-generating function. By using this approach to propagate uncertainties through STTs, we effectively also mitigate potential issues arising from distribution tails extending beyond the radius of convergence of the expanded flow, as we ensure that probability density functions have a compact domain fully contained within this radius.
We test the proposed method leveraging map-inversion techniques \cite{berz1988differential} and propagating uncertainties forward or backward at predefined events, such as crossing a surface of section or similar. 

\section{Methods}
\subsection{High-Order Taylor Maps}
\subsubsection{Background}
\label{sec:background}
In this section, we briefly discuss the use of high-order Taylor expansions to represent perturbations around a nominal trajectory.
These formulations have been discussed profusely in past works~\cite{berz1988differential, berz1998verified, jorba2005software, jones2013nonlinear, park2006nonlinear}, often using different terminologies. 
The very same quantities (i.e., the coefficient of a multivariate Taylor polynomial) have been referred to with different names 
such as differential algebraic expansion of the flow, jet transport of the derivatives, state transition tensors, flow expansion via the algebra of truncated Taylor polynomials, and more.
Effectively, all these methods provide some way to compute the Taylor coefficients of the flow expansion and related quantities. We will here embrace and use the \lq\lq state transition tensors\rq\rq\ terminology, as the term state transition matrix has arguably a most immediate understanding in the context of aerospace engineering.

Consider a generic dynamical system described by the following set of $n$ ODEs:

\begin{equation}
    \dot{\pmb{x}}=\pmb{f}(\pmb{x},\pmb{q})\text{,}
    \label{eq:ode_dynamical_system}
\end{equation}
where $\pmb{f}\in \mathbb{R}^n$ represents the system dynamics, $\pmb{x}\in \mathbb{R}^n$ is the state vector and $\pmb{q}\in \mathbb{R}^m$ a vector of parameters. We write the deviations of the state along a trajectory solution of Eq.~\eqref{eq:ode_dynamical_system} using the Taylor expansion of the flow $\pmb{\phi}(t;\pmb{x}_0,\pmb{q})$  which we also indicate with $\pmb{x}_f$:
\begin{equation}
    \delta x^i_f= \mathcal{P}_{i}^k(\delta \pmb{x}_0, \delta \pmb{q})+\mathcal{O}(k)
    \label{eq:taylor_expansion}
\text{,}
\end{equation}
where $\delta x_f^i$ is the $i$-th component of the state deviation at the final time, while $\mathcal{P}_{i}^k$ is the $k$-the order Taylor polynomial for the $i$-th component of the final state deviations (each component of the state is described by its own polynomial). The symbol $\mathcal{O}(k)$ indicates the terms of the series that have an order higher than $k$, which will be ignored when using the truncated Taylor expansion. 
Furthermore, we use $\delta \pmb{x}_0=\pmb{x}_0-\overline{\pmb{x}}_0$ to indicate perturbations of the nominal initial state $\overline{\pmb{x}}_0$, and $\delta \pmb{q}=\pmb{q}-\overline{\pmb{q}}$ to indicate the deviations with respect to the nominal parameters $\overline{\pmb{q}}$. 

While the above polynomials can be equivalently expressed in various notations, we here use the compact multi-index notation~\cite{reed1980methods}:
\begin{equation}
    \delta x_f^i\approx \mathcal{P}_i^k(\delta \pmb{z})=\sum_{|\alpha|=1}^k \dfrac{1}{\alpha!}(\partial^\alpha x_{f}^i)\bigg|_{(\overline{\pmb{x}}_0, \overline{\pmb{q}})}\delta \pmb{z}^\alpha\text{,}
    \label{eq:taylor_polynomials_multi_index}
\end{equation}
where $k$ is the order of the Taylor polynomial expansion and $\alpha=(\alpha_1,\dots,\alpha_n)$ is the $n$-tuple of non-negative integers associated with each component of the vector $\pmb{x}_f\in\mathbb{R}^n$, where its factorial is defined as $\alpha!=\alpha_1!\cdot \dots \cdot \alpha_n!$. Furthermore, $|\alpha|=\sum_{j=0}^M\alpha_j$ must be taken over all possible combinations of $\alpha_j\in\mathbb{N}$. We refer to $\delta \pmb{z}=[\delta\pmb{x}_0^T, \delta \pmb{q}^T]^T$ as the vector combining the initial state perturbations and parameters deviations.

In summary, Eq.~\eqref{eq:taylor_polynomials_multi_index}, which we also refer to as a high-order Taylor map, provides a way to map initial perturbations around a nominal trajectory, at future times.

The largest norm of the perturbation $\pmb{\delta z}$ within which the series converges is referred to as the radius of convergence. 
When dealing with expansions such as that in Eq.~\eqref{eq:taylor_polynomials_multi_index}, it is important to estimate the radius of convergence as well as the truncation error. 
We use an estimate of the radius of convergence to check that all uncertainties are within the convergence radius (a necessary condition that allows the use of the expansions for their propagation). 
As for the truncation error, we compute it exactly and report its value to cross-validate the specific computations presented here. Note that, in principle, the truncation error can also be bounded rigorously, e.g. using Taylor models~\cite{makino2003taylor}, but not necessarily in a sufficiently tight enclosure.

\subsubsection{Radius of Convergence}
To compare how well the Taylor expansions can reconstruct the actual nonlinear dynamics, we can, as an initial analysis of the convergence radius, use results from basic calculus that study the convergence of power series. In this context, two different criteria are among the most popular ones: the ratio test (i.e., D'Alembert's criterion)~\cite{bromwich2005introduction} and the Cauchy-Hadamard theorem~\cite{cauchy1821cours,hadamard1892essai}. On the one hand, the former computes the convergence radius as:
\begin{equation}
    R_{c}= \lim_{k\rightarrow \infty}\dfrac{||b_k||}{||b_{k+1}||}\text{,}
    \label{eq:dalembert}
\end{equation}
where $b_k$ are the matrices resulting from unfolding the STT associated with the $k$-th Taylor term and $||.||$ indicates some norm, for example the 2-norm~\cite{izzo2020stability}. On the other hand, the Cauchy-Hadamard convergence radius, for multivariate cases, can be written as~\cite{bekbaev2013radius}:
\begin{equation}
R_{c}=1/\lim_{k\rightarrow\infty}\textrm{sup}_{|\alpha|=k}|a_{\alpha}\big( \alpha!/k!\big)^{1/2}|^{1/k}
\text{,}
    \label{eq:cauchy_hadamard}
\end{equation}
where $a_{\alpha}$ is the $\alpha$-coefficient of the Taylor series, with $\alpha$ being the $n$-tuple already introduced in Eq.~\eqref{eq:taylor_polynomials_multi_index}.

In this work, we performed experiments with both criteria, observing that at truncation orders up to five, and for the orbital dynamics cases studied, the Cauchy-Hadamard test produces more reliable estimates of the radius of convergence. Furthermore, it has been found that the Cauchy-Hadamard test displays a more reliable behavior as the order increases. In contrast, the ratio test provided more oscillating behaviors, making it more difficult to interpret the accuracy of the Taylor polynomial series as the order is increased. 
In Sec.~\ref{sec:cr3bp_experiments}, we show numerical experiments on the quality of the convergence radius estimated by the Cauchy-Hadamard criteria.

\subsubsection{Computing the Taylor Coefficients}
Before delving into the connection between the high order Taylor maps introduced and the propagation of statistical moments, it is essential to clarify some practical details regarding the computation of this Taylor expansion. In literature, two main methods exist for the computation of the partial derivatives of the flow with respect to initial conditions and parameters: the application of some flavor of automatic differentiation techniques through a numerical integrator, which can be done via modern machine learning tools~\cite{paszke2019pytorch, jax2018github} or via the algebra of truncated Taylor polynomials~\cite{makino2006cosy, massari2018differential, izzo2018audi}, or the direct numerical integration of the high-order variational equation~\cite{park2006nonlinear, jorba2005software}. 
In \cite{gimeno2023numerical}, the authors show the mathematical equivalence of the variational equation approach to high-order automatic differentiation, for the same numerical integrator scheme. Hence, given a numerical integrator scheme, the same polynomial expansions will be obtained. While this is true mathematically, the implementation details can play a big difference in the computational time needed to compute these quantities practically. 
In previous work~\cite{valli2013nonlinear}, for instance, the use of differential algebra (in that context a forward mode automated differentiation technique) to compute the STTs was reported to be orders of magnitude more efficient than the variational approach.
In contrast to that finding, we use the variational approach to compute STTs for all the cases presented in Sec.~\ref{sec:numerical_results} and found that it is as competitive as differential algebraic techniques while allowing for the seamless use of Taylor integrators and the resulting reliable event detection machinery described in Ref. \cite{biscani_heyoka, biscani2022reliable}.
This finding sharply contrasts with the conclusions drawn by Valli et al.\cite{valli2013nonlinear}, likely due to the assumption they made on the number of variational equations needed to compute the needed tensors. In Valli's work, this is presumed to grow linearly with $N = \sum_{q=1}^{k+1} n^q$ where $n$ is the state dimension and $k$ is the expansion order. The same assumption was also shared by earlier studies, such as~\cite{park2006nonlinear}. However, this scaling fails to apply to the variational equations we derive automatically using the \textit{heyoka} variational ODE tool\footnote{\url{https://bluescarni.github.io/heyoka/}, accessed October 2024.}~\cite{biscani_heyoka}. In this case, the number of variational equations $N$ incorporates all symmetries emerging from Schwarz's theorem, thereby resulting in a significantly reduced count~\cite{izzo2017differentiable}:%griewank2000evaluating
$$ 
N = n \sum_{q=1}^{k}\left(\begin{array}{c}q+n-1\\q\end{array}\right).
$$
The final number is orders of magnitude smaller than previously assumed as the order increases. For example, in the case of $n=6$ equations, varying $k$ up to the fifth order we have [42, 168, 504, 1260, 2772] variational equations, while previous assumptions expected this number to grow as [42, 258, 1554, 9330, 55986].

\subsection{Probability Distributions}
\subsubsection{Fokker-Planck equation.}
Consider a dynamical system where both the state of the system and the parameters of the dynamics are random vectors, described by (piecewise-)continuous probability density functions. 
Denote with $p(\pmb{x},\pmb{q},t)$ the combined probability distribution of the state at time $t$ and parameters. At time $t=0$, this will be some known function $p_0(\pmb{x},\pmb{q})$ describing our initial uncertainties. Assuming no diffusion acting on the dynamics, the probability density function $p$ will evolve through our dynamics $\pmb{f}$ according to the Fokker-Planck equation:
\begin{equation}
    \dfrac{\partial p(\pmb{x},\pmb{q},t)}{\partial t}=-\sum_{i=1}^n\dfrac{\partial}{\partial x_i}\big[p(\pmb{x},\pmb{q},t)f_i \big]
\end{equation}
To compute the probability that random variable vectors take on a value in any given interval, one first needs to solve the above equation, and then integrate the PDF over that interval. 
While simplifications arise when the dynamical system is Hamiltonian because, according to Liouville's theorem, the phase-space probability density function remains constant along the system's trajectories, the above equation is a partial differential equation that becomes particularly cumbersome to solve, especially for nonlinear dynamics and high-dimensional problems. Instead, we work with only the first moments of the distribution: these are generally descriptive of the PDF and do not require the full solution of the Fokker-Planck equation. 
In the next section, we will introduce several definitions associated with these quantities.

\subsubsection{Expectation Operator of Continuous Random Variables}
\label{sec:statistical_moments}
Assuming that the state is $n$-dimensional and the parameters are $m$-dimensional, then, the expectation of each component of the state can be computed by solving the multi-dimensional integral:
\begin{equation}
    \EX[x_i]=\int_{\mathbb{R}^n}\int_{\mathbb{R}^m}x_ip(\pmb{x},\pmb{q})d\pmb{x}d\pmb{q}\text{.}
    \label{eq:expectation}
\end{equation}

Several estimators exist to evaluate the high-dimensional integrals in Eq. \eqref{eq:expectation}, including Monte Carlo methods and Gaussian quadrature, among others~\cite{jazwinski2007stochastic}. 
As discussed in the next sections, we will leverage a combination of moment generating functions and Taylor polynomials to compute expectations of the random variables efficiently without having to calculate the high-dimensional %complex 
integrals of Eq. \eqref{eq:expectation}.
This enables us to quickly calculate the moments of a distribution owing to their mathematical relationship with the expected values of continuous random variables.

\subsubsection{Statistical Moments}
Moments about the mean (also known as central moments) are often preferred to the ones about the origin because they carry more information about the shape and spread of the distribution, rather than its location (since they are computed w.r.t. the mean). For a multivariate case, their computation can be performed as follows:
{\small
\begin{equation}
\begin{cases}
    m_1&=[\EX[x_1],\dots, \EX[x_n]]\\
    m_{2,ij}&=\EX[(x_i-\EX[x_i])(x_j-\EX[x_j])]\\
    \dots\\
    m_{k,ij\dots l}&=\EX[(x_i-\EX[x_i])(x_j-\EX[x_j])\dots (x_l-\EX[x_l])]\text{.}
\end{cases}    
\label{eq:first_k_moments}
\end{equation}
}
For the case of the covariance (i.e., the second equation in Eq.~\eqref{eq:first_k_moments}), we can further develop each term as:
\begin{align}
    \begin{split}
    m_{2,ij}&=\EX[(x_i-\EX[x_i])(x_j-\EX[x_j])]\\
    &=\EX[x_ix_j]-\EX[x_i]\EX[x_j]-\EX[x_i]\EX[x_j]+\EX[x_i]\EX[x_j]\\
    &=\EX[x_ix_j]-\EX[x_i]\EX[x_j]
\text{.}
    \end{split}
    \label{eq:second_moment_expression_simplified}
\end{align}
For the third central moment, instead, we would have:
\begin{align}
    \begin{split}
    m_{3,ijk}&=\EX[(x_i-\EX[x_i])(x_j-\EX[x_j])(x_k-\EX[x_k])]\\    &=\EX[x_ix_jx_k]+2\EX[x_i]\EX[x_j]\EX[x_k]+\\
    &\ \ -\EX[x_i]\EX[x_jx_k]-\EX[x_j]\EX[x_ix_k]-\EX[x_k]\EX[x_ix_j]
\text{.}
    \end{split}
    \label{eq:third_central_moment_formula}
\end{align}

Similar formulas can also be derived for higher central moments. The key message is that moments about the mean can be written in terms of moments about the origin.
In most applications, only the first two moments (mean and covariance) are considered. Hence, having a method that accurately represents the first two moments often suffices. However, there are benefits in having access to higher-order moments, especially if these can be calculated in a numerically efficient way as proposed in the following Sections. Among the other use cases listed in ~\cite {kendall1946advanced}, some of these benefits include:
\begin{enumerate}
    \item Statistical analysis: Higher moments can give insights into the actual shape of the distributions (e.g. its symmetry and its fat-tailedness), which would not be available from the first two moments. This can also be used to prove/disprove the normality of a PDF, for instance by monitoring whether odd moments about the mean are zero (which is the case for symmetric distributions, like the Gaussian) or not. 
    \item Probability density function estimation: Using the first few statistical moments, one can provide an approximated estimate of the probability density function, that more closely resembles the true PDF. In practice, this implies finding an approximate solution to the "moment problem"~\cite{shohat1950problem}. Although mathematically it can be demonstrated that for distributions with infinite support, this problem can be indeterminate, in practical scenarios, several techniques (e.g. maximum entropy estimation) can be used to approximate the PDF. In most of those approaches, having higher moments results in a more accurate estimation of the probability density function~\cite{wakefield2023moment, shohat1950problem, akhiezer2020classical}.
    \item Probability bounds: Finally, higher moments can be used to have tighter bounds on the probability of the random variables. Examples include Markov inequality and Chebyshev inequality~\cite{ghosh2002probability}.
\end{enumerate}
\subsubsection{Moment Generating Functions}
Moment generating functions (MGFs) are mathematical functions often employed in probability theory and statistics to compute moments of a probability density function. In particular, for a generic $n$-dimensional vector $\pmb{x}$, these functions are defined as:
\begin{equation}
    M_{\pmb{x}}(\pmb{t})=\EX[e^{\pmb{t}^T\pmb{x}}]\text{.}
\end{equation}
As suggested by their name, they are strictly connected to statistical moments, in particular, if these functions exist in an open interval around $\pmb{t}=\pmb{0}$, then the following relationship holds:
\begin{equation}
    \EX[x_1^{k_1}\dots x_n^{k_n}]=\dfrac{\partial^k}{\partial t_{1}^{k_1}\dots \partial t_{n}^{k_n}}M_{\pmb{x}}(\pmb{t})\bigg|_{\pmb{t}=0}\text{.}
    \label{eq:moments_about_origin_and_MGF}
\end{equation}

Moment generating functions are known for a wide class of PDFs, including univariate and multivariate Gaussian, uniform distributions, Bernouilli, binomial, exponential, beta, gamma, Chi-squared, and more. In Tab.~\ref{tab:MGF}, we present some MGFs for a variety of PDFs. 
Moreover, we recall that for an $n$-dimensional random variable, where each component is independently randomly distributed, then the moment generating function for the $n$-dimensional random vector can be written as the product of their moment generating functions:

\begin{equation}
    M_{\pmb{x}}(\pmb{t})=\prod_{i=1}^n M_{x_i}(t_i)\text{.}
    \label{eq:multivariate_MGF_independent}
\end{equation}
\begin{table*}[!tb]
    \centering
    \scriptsize
    \caption{Moment Generating Functions of Various Distributions.}
    \begin{tabular}{|l|l|l|}
        \hline
        \textbf{Distribution Name} & \textbf{Probability Density Function} & \textbf{Moment Generating Function (MGF)} \\
        \hline
        Degenerate & $\textrm{Pr}(X = c) = 1$ & $M_X(t) = e^{t c}$ \\
        \hline
        Normal & $p(x) = \frac{1}{\sqrt{2 \pi \sigma^2}} \exp\left(-\frac{(x-\mu)^2}{2\sigma^2}\right)$ & $M_X(t) = \exp\left( \mu t + \frac{1}{2} \sigma^2 t^2 \right)$ \\
        \hline
        Multivariate Normal & $p(\pmb{x}) = \frac{1}{(2\pi)^{k/2} |\boldsymbol{\Sigma}|^{1/2}} \exp\left(-\frac{1}{2}(\pmb{x} - \pmb{\mu})^\top \pmb{\Sigma}^{-1} (\pmb{x} - \pmb{\mu})\right)$ & $M_{\pmb{X}}(\pmb{t}) = \exp\left( \pmb{t}^\top \pmb{\mu} + \frac{1}{2} \pmb{t}^\top \pmb{\Sigma} \pmb{t} \right)$ \\
        \hline
        Uniform & $p(x) = \frac{1}{b-a}, \quad a \le x \le b$ & $M_X(t) = \frac{e^{tb} - e^{ta}}{t(b-a)}$ \\
        \hline
        Chi-Squared & $p(x) = \frac{1}{2^{k/2} \Gamma(k/2)} x^{k/2 - 1} e^{-x/2}, \quad x > 0$ & $M_X(t) = (1 - 2t)^{-k/2}, \quad t < \frac{1}{2}$ \\
        \hline
        Gamma & $p(x) = \frac{\beta^\alpha}{\Gamma(\alpha)} x^{\alpha - 1} e^{-\beta x}, \quad x > 0$ & $M_X(t) = \left(1 - \frac{t}{\beta}\right)^{-\alpha}, \quad t < \beta$ \\
        \hline
        Geometric & $\textrm{Pr}(X = k) = (1 - \alpha)^k \alpha, \quad k = 0, 1, 2, \ldots$ & $M_X(t) = \frac{\alpha e^t}{1 - (1-\alpha)e^t}, \quad t < -\ln(1-\alpha)$ \\
        \hline
        Bernoulli & $\textrm{Pr}(X = 1) = \alpha, \quad Pr(X = 0) = 1 - \alpha$ & $M_X(t) = 1 - \alpha + \alpha e^t$ \\
        \hline
        Poisson & $\textrm{Pr}(X = k) = \frac{\lambda^k e^{-\lambda}}{k!}, \quad k = 0, 1, 2, \ldots$ & $M_X(t) = \exp\left( \lambda (e^t - 1) \right)$ \\
        \hline
        Binomial & $\textrm{Pr}(X = k) = \binom{n}{k} \alpha^k (1 - \alpha)^{n - k}, \quad k = 0, 1, 2, \ldots, n$ & $M_X(t) = (1 - \alpha + pe^t)^n$ \\
        \hline
        Exponential & $p(x) = \lambda e^{-\lambda x}, \quad x \ge 0$ & $M_X(t) = \frac{\lambda}{\lambda - t}, \quad t < \lambda$ \\
        \hline
    \end{tabular}
    \label{tab:MGF}
\end{table*}
Due to their relevance for practical applications, we will discuss in detail the case for multivariate normal and multivariate uniform distributions. 

For a multivariate normal distribution, the MGF can be written as:

\begin{equation}
    M_{\pmb{x}}(\pmb{t})=\textrm{exp}(\pmb{\mu}^T\pmb{t}+\dfrac{1}{2}\pmb{t}^T \Sigma \ \pmb{t})
    \text{,}
    \label{eq:moment_generating_function}
\end{equation}
where $\pmb{x}\in \mathbb{R}^n$, $\pmb{t}\in \mathbb{R}^n$. Furthermore, $\pmb{\mu}\in \mathbb{R}^n$ is the first moment (i.e., mean vector), and $\Sigma\in \mathbb{R}^{n\times n}$ is the second moment (i.e., covariance matrix). %Isserlis' theorem provides a formula to compute high-order central moments for a multivariate Gaussian, only as a function of the second moment~\cite{isserlis1918formula}:
By taking the partial derivatives in Eq.~\eqref{eq:moments_about_origin_and_MGF}, it can be shown that for a zero-mean multivariate normal random vector, it holds:
\begin{equation}
    \EX[x_1^{k_1}\dots x_n^{k_n}]=\sum_{p}\prod_{\{i,j\}\in p}\sigma_{ij}\text{,}
\end{equation}
where the sum is over all allocations of the set $\{1,\dots,k\}$, with $k=\sum_ik_i$, into $k/2$ unordered pairs. While $\sigma_{ij}$ indicates the covariance of $x_i$ and $x_j$. This formula holds if $k$ is even, while if $k$ is odd, the $k$-th order central moment is zero. This result is also known as Isserlis' theorem or Wick's theorem~\cite{isserlis1918formula}. 

As for uniform probability density functions, the MGF of the univariate case can be written as:
\begin{equation}
M_{x_i}(t_i)=
\begin{cases}
    \dfrac{e^{t_ib_i}-e^{t_ia_i}}{t_i(b_i-a_i)} \hspace{2cm}& t_i\neq 0\\
    1  &\textrm{otherwise}\text{.}
\end{cases}    
\end{equation}
Note that the discontinuity at $t=0$ is removable (the two one-sided limits are equal), hence value and derivatives are well-defined. In fact, by expanding the exponential terms with a Mclaurin series, the MGF can be written in the form:
\begin{equation}
    M_{x_i}(t_i)=\sum_{k=0}^{\infty}\dfrac{(b_i^{k+1}-a_i^{k+1})}{(b_i-a_i)(k+1)}\dfrac{t_i^k}{k!}\text{,}
    \label{eq:MGF_unifrom_univariate}
\end{equation}
which is free of any singularity. We can now take the partial derivatives of Eq.~\eqref{eq:MGF_unifrom_univariate} and find:

\begin{equation}
    \EX[x_i^k]=\dfrac{b_i^{k+1}-a_i^{k+1}}{(b_i-a_i)(k+1)}\text{.}
    \label{eq:moment_n_univariate_uniform}
\end{equation}
The multivariate uniform case is then easily constructed by combining Eq.~\eqref{eq:multivariate_MGF_independent} and Eq.~\eqref{eq:moment_n_univariate_uniform}:
\begin{equation}
    \EX[x_1^{k_1}\dots x_n^{k_n}]=\prod_{i=1}^n \dfrac{b_i^{k_i+1}-a_i^{k_i+1}}{(b_i-a_i)(k_i+1)}\text{.}
\end{equation}

For a given probability density function (provided it admits an MGF), the high-order moments can be computed one-off using a symbolic manipulator able to automatically differentiate expressions efficiently up to high orders. In our case, we utilize \textit{heyoka}, either through the partial derivatives of the MGF or through recursive formulas where available (e.g., for the uniform or Gaussian distribution). This allows the computational cost to be decoupled from a specific dynamical system, enabling these terms to be computed one-off, stored, and reused.

\subsection{Nonlinear Mapping of Statistical Moments of Perturbations}
\label{sec:nonlinear_mapping_of_statistical_moments_of_perturbations}

\subsubsection{General Formulation}
\label{sec:general_formulation}
Due to the linearity of the expectation operator, by taking the expectation of perturbations around the nominal solution flow, shown in Eq.~\eqref{eq:taylor_polynomials_multi_index}, we obtain:

\begin{equation}
    \EX[\delta x_f^i]\approx \EX[\mathcal{P}_i^k(\delta \pmb{z})]=\sum_{|\alpha|=1}^k \dfrac{1}{\alpha!}(\partial^\alpha x_f^i)\bigg|_{(\overline{\pmb{x}}_0, \overline{\pmb{q}})}\EX[\delta \pmb{z}^\alpha]\text{.}
    \label{eq:expectation_flow_perturbations}
\end{equation}

This equation also serves as a starting point in other studies, such as those by Park et al. (2006) and Valli et al. (2013)~\cite{park2006nonlinear, valli2013nonlinear}, where it is assumed that perturbations in initial conditions and parameters follow a Gaussian distribution. This assumption allows the application of Isserlis' theorem to determine higher moments based on the first two moments. However, the approach is limiting because it only accommodates Gaussian univariate or multivariate distributions. In this work, we introduce a more general method that can handle arbitrary probability density functions (PDFs). A notable application of this method is in the context of uncertain gravitational fields around small celestial bodies. In such cases, it is appropriate to model the uncertainties in gravitational coefficients with uniform distributions, rather than the Gaussian distributions commonly assumed in the literature, such as in the work of Panicucci et al. (2020)~\cite{panicucci2020uncertainties}.

To remove the Gaussian hypothesis from the distribution of $\delta \pmb{z}$, instead of Isserlis' theorem, we leverage Eq.~\eqref{eq:moments_about_origin_and_MGF} to compute the statistical moments of the initial state and parameters. The moments are later propagated forward or backward in time via Eq. ~\eqref{eq:expectation_flow_perturbations}. A similar scheme also works for higher moments. For instance, the second moment elements can be written as:
{\small
\begin{align}
\begin{split}
m_{2,ij}&=\EX[(\delta x_f^i-\EX[\delta x_f^i])(\delta x_f^j-\EX[\delta x_f^j])]\approx \\
&\sum_{|\alpha|=1}^k \sum_{|\beta|=0}^k \dfrac{1}{\alpha!}\dfrac{1}{\beta!}(\partial^\alpha x_f^i)\bigg|_{(\overline{\pmb{x}}_0, \overline{\pmb{q}})}(\partial^\beta x_f^j)\bigg|_{(\overline{\pmb{x}}_0, \overline{\pmb{q}})}\EX[\delta \pmb{z}^\alpha\delta \pmb{z}^\beta]+\\
&-\EX[\delta x_f^i]\EX[\delta x_f^j]
%\label{eq:}
\end{split}
\end{align}}
where $|\beta|=\sum_{s=0}^M\beta_s$ must also be taken over all possible combinations of $\beta_s \in \mathbb{N}$. Note that we take advantage of the expression derived in Eq.~\eqref{eq:second_moment_expression_simplified} to express the second moment, and that similar expressions can be derived for higher ones. These expressions will involve the computations of the higher-order moment about the origin (see, for example, Eq.~\eqref{eq:third_central_moment_formula}), plus a series of coefficients that can all be computed from the lower-order moments.

In this way, one can compute only once, independently of the dynamical system of interest, expressions for these expectations, which can then just be evaluated at different parameters of the distribution, depending on the case at hand. Hence, the computational cost of the method for propagating the statistical moments is therefore reduced to the sole expansion of the flow around its nominal trajectory.

As a final note, we highlight that, with the proposed method, the $m$-th order moment will be composed of terms that are generated by the product of $m$ different polynomials, each expanded up to order $k$. This will produce, as a result, polynomials of increasing degree (i.e., by multiplying two polynomials of order $k$, one obtains a polynomial of order $k^2$). If one were to work with the algebra of truncated polynomials, then truncating at order $k$ would cause all higher-order contributions to disappear, which would necessarily cause a degradation of accuracy. Hence, in these cases, it is important to retain higher-order terms.

\subsubsection{Mapping Statistical Moments via High-order Maps}
The formalism introduced above can be applied to map statistical moments of any given PDF through polynomial expressions, provided that the PDF admits a moment generating function. 
When using STTs these polynomial expressions represent the flow of our dynamical system, but we can also obtain polynomials to represent the state of an arbitrary event equation (e.g. this could be the mapping between surfaces of section crossings in a dynamical system). 
The resulting Event Transition Tensors (ETTs) are computed as follows.

We introduce the event manifold, implicitly defined as $e(\pmb{x}, \pmb{q})=0$. At the event trigger time, $t^*$, we have:
\begin{equation}
e(\pmb{\phi}(t^*;\pmb{x}_0,\pmb{q}),\pmb{q})=0\text{,} 
\end{equation}
with $\pmb{\phi}$ being the flow of the dynamical system in Eq.~\eqref{eq:ode_dynamical_system}.

A particular event case, which is quite relevant for astrodynamics application, are surface of sections. These are $n-1$-dimensional surfaces embedded in an $n$-dimensional phase space. States on the surface of section are mapped forward and backward in time via discrete Poincar\'{e} maps that enable the qualitative study of dynamical systems~\cite{poincare1893methodes}.

Mapping statistical moments up to an event in a traditional Monte Carlo-based approach would involve sampling the parameters and initial state uncertainties and stopping the numerical integration at the event  (e.g. when each particle meets the event specified in the integrator). 

Instead, having access to the Taylor expansion of the state at future times enables \textcolor{red}{\st{us}} to construct an accurate representation of the nonlinear event map by means of inversion techniques~\cite{hawkes1999modern, armellin2021collision, armellin2010asteroid}. 
Figure ~\ref{fig:schematic_illustration_poincare_map_up} illustrates two examples in which the moments are mapped either at an event represented by a surface of section (in this case a plane) of a periodic orbit (on the left), or for the case in which the event is constituted by the whole surface of a target asteroid body (on the right). 

\begin{figure}
    % Legend
    \begin{tikzpicture}
        \begin{scope}[shift={(-4, 2)}, scale=0.6] % Adjust the shift and scale to position the legend
            % Poincare map
            \draw[blue,fill=blue!10,opacity=0.5] (-0.3,0) rectangle (0.4,0.2);
            \node[right] at (0.5, 0.1) {Surface of Section};

            % Initial covariance
            \draw[green] (0,-0.5) ellipse (0.2 and 0.1);
            \node[right] at (0.5, -0.4) {Initial covariance};

            % Final covariance
            \draw[red] (0,-1) ellipse (0.2 and 0.1);
            \node[right] at (0.5, -0.9) {Final covariance};
        \end{scope}
    \end{tikzpicture}

    % First TikZ picture: Poincaré map illustration with spacecraft
    \begin{minipage}[t]{0.48\linewidth} % Adjust width as necessary
        \centering
        \begin{tikzpicture}[scale=1]
            % Define the closed orbit (circle for simplicity)
            \draw[thick] (0,0) circle (1);
            % Define the point on the orbit (upper part)
            \fill[red] (0,1) circle (0.05);
            % Draw the orthogonal plane (as a rotated and inclined square) at the point
            \begin{scope}
                \clip (-2,-1) rectangle (2,3);
                \draw[thick,blue,fill=blue!10,opacity=0.5,rotate around={60:(0,1)}] (0,1) ++(-0.75,-0.75) rectangle ++(1.5,1.5);
            \end{scope}
            % Draw the smaller covariance ellipse
            \draw[thick,green,rotate around={60:(0,1)}] (0,1) ellipse (0.4 and 0.2);
            % Draw the larger covariance ellipse
            \draw[thick,red,rotate around={60:(0,1)}] (0,1) ellipse (0.55 and 0.275);
            % Draw the spacecraft centered on the circular orbit
            \begin{scope}[shift={(-1,-0.27)}, rotate=-30, scale=0.4]
                % Draw the cube body of the spacecraft, centered on the trajectory
                \draw[thick] (-0.2, -0.2) rectangle (0.2, 0.2);
                % Draw the solar panels as lines with rectangles on top
                \draw[thick] (-0.2, 0) -- (-0.8, 0);
                \draw[thick] (-0.8, -0.1) rectangle (-0.6, 0.1);
                \draw[thick] (0.2, 0) -- (0.8, 0);
                \draw[thick] (0.6, -0.1) rectangle (0.8, 0.1);
            \end{scope}
        \end{tikzpicture}
    \end{minipage}
    \hfill
    % Second TikZ picture: Sphere and spacecraft
    \begin{minipage}[t]{0.48\linewidth} % Adjust width as necessary
        \centering
        \begin{tikzpicture}[scale=0.49]
            % Draw the sphere
            \shadedraw[ball color=gray] (0,0) circle (2);
            % Draw the point far away
            \filldraw (6, 4) circle (0.1) node[above right] {};
            % Draw the curved trajectory
            \draw[thick, dashed, ->] (6, 4) .. controls (3, 6) and (1, 5) .. (0, 0.5);
            % Draw the small green ellipsoid around the point
            \begin{scope}[shift={(6,4)},rotate=30]
                \draw[green, thick] (0, 0) ellipse (0.5 and 0.3);
            \end{scope}
            % Draw the red ellipsoid on the surface of the sphere, slightly above the center
            \begin{scope}
                \clip (0,0) circle (2);
                \begin{scope}[shift={(0,0.5)},rotate=0]
                    \draw[red, thick] (0, 0) ellipse (0.7 and 0.4);
                \end{scope}
                % Draw a black dot at the center of the red ellipsoid
                \filldraw[black] (0, 0.5) circle (0.05);
            \end{scope}
            % Draw the spacecraft in the middle of the trajectory
            \begin{scope}[shift={(3., 5.)}, rotate=-30]
                % Draw the cube body of the spacecraft, centered on the trajectory
                \draw[thick] (-0.2, -0.2) rectangle (0.2, 0.2);
                % Draw the solar panels as lines with rectangles on top
                \draw[thick] (-0.2, 0) -- (-0.8, 0);
                \draw[thick] (-0.8, -0.1) rectangle (-0.6, 0.1);
                \draw[thick] (0.2, 0) -- (0.8, 0);
                \draw[thick] (0.6, -0.1) rectangle (0.8, 0.1);
            \end{scope}
        \end{tikzpicture}    
    \end{minipage}
    
    \caption{Schematic examples of mapping of moments onto a Poincaré map (left) and onto the surface of an asteroid (right).}
    \label{fig:schematic_illustration_poincare_map_up}
\end{figure}

Without loss of generality, we consider the case in which the surface of the section is defined as the return map for one of the coordinates, meaning that it verifies $x_f^j-\overline{x}_0^j=0$, for a given $j$. As already discussed in previous work (e.g., ~\cite{fu2024high} and references therein), by expanding the final state perturbation as done in Eq.~\eqref{eq:taylor_polynomials_multi_index} and including the integration time as one of the parameters, the map can be partially inverted, and the Taylor polynomial of the integration time corrections needed to ensure that any perturbation around the initial state and parameter returns to the surface of section can be found:
\begin{equation}
    \delta t^*=\delta t^*(\delta \pmb{x}_0, x_f^j-x_0^j=0, \delta \pmb{q})\text{,}
\end{equation}
where $\delta t^*$ indicates the period correction that enables the return at the surface of section. Moreover, by substituting this Taylor polynomial into Eq.~\eqref{eq:taylor_polynomials_multi_index}, one can obtain an expression for the reduced state (i.e., the $n-1$ state coordinates at the surface of section): this will also be a Taylor polynomial of order $k$. Hence, at the only additional cost of partially inverting the map and making a few substitutions, and using the same approach detailed above, one can directly propagate the statistical moments at the event manifold. 

In Sec.~\ref{sec:numerical_results}, we will first tackle two cases related to the two-body problem (unperturbed and J2-perturbed), in which the moments are mapped at future times (i.e., stroboscopic mappings), and then, we will also introduce and showcase results for the moments' propagation via Poincaré maps obtained in the circular restricted three-body problem (CR3BP) of the Earth-Moon and Saturn-Enceladus systems.

\section{Numerical Results}
\label{sec:numerical_results}
\subsection{Two-Body Problem}
\subsubsection{Unperturbed Case}
\label{sec:simple_case_two_body}
As a first experiment to test the proposed scheme for the propagation of statistical moments of non-Gaussian initial conditions, we will cover a simple application in a two-body problem given by
\begin{align*}
    \dot{\pmb{r}}&=\pmb{v}\\
    \dot{\pmb{v}}&=-\dfrac{\mu}{|\pmb{r}|^3}\pmb{r}
\end{align*}
where $\pmb{x}=[\pmb{r}^T,\pmb{v}^T]^T$ is the state of the spacecraft and $\mu$ is its gravitational parameter.
This case serves as a toy model that can be well understood to illustrate the usage of the method and its usefulness. We begin by considering a planar Earth circular orbit and by normalizing the time and length units of the system such that both the semi-major axis of the satellite and gravitational parameter are equal to 1. Furthermore, it is assumed that the spacecraft's initial velocity vector is sampled from a degenerate probability density function (i.e., the Dirac-delta function), centered in $\pmb{v}_0=[0,1,0]^T$, while its initial position vector is uncertain and described by a uniform distribution with mean: $\overline{\pmb{r}}_0=[1,0,0]^T$ and upper and lower bounds of $\pm10^{-2}$ from the mean. The central body has an uncertain gravitational parameter described by a uniform distribution with mean $\overline{\mu} = 1$ and upper and lower bounds of $\pm0.5\times10^{-2}$. 
The spacecraft is then propagated for $2\pi$ units of time (i.e., a full orbit). Since uncertain parameters and initial conditions are propagated in a coupled dynamical system, the overall state will be uncertain, as a result. 

We compare the statistics reconstructed via the moment generating function method we propose, and the sample statistics reconstructed via Monte Carlo propagation. In particular, we compute the Monte Carlo statistics by sampling 10 million points from the uncertain PDF of the state and parameter and propagating each of them at the final integration time. This is compared against the method described in Sec.~\ref{sec:nonlinear_mapping_of_statistical_moments_of_perturbations}, which only requires a single Taylor expansion around the nominal state, combined with the cost of computing the expectation over the initial probability density functions (which, however, can be performed only once, offline and independently of the specific dynamics). While depending on the chosen order of the expansion, the computational cost can increase substantially: in our case, we find that for all the tested orders (i.e., up to order 4), the polynomial map can be very quickly constructed, at orders of magnitude less computation time than the Monte Carlo approach. 

\begin{figure*}[tbh!]
  \centering
  \includegraphics[width=1.\columnwidth]{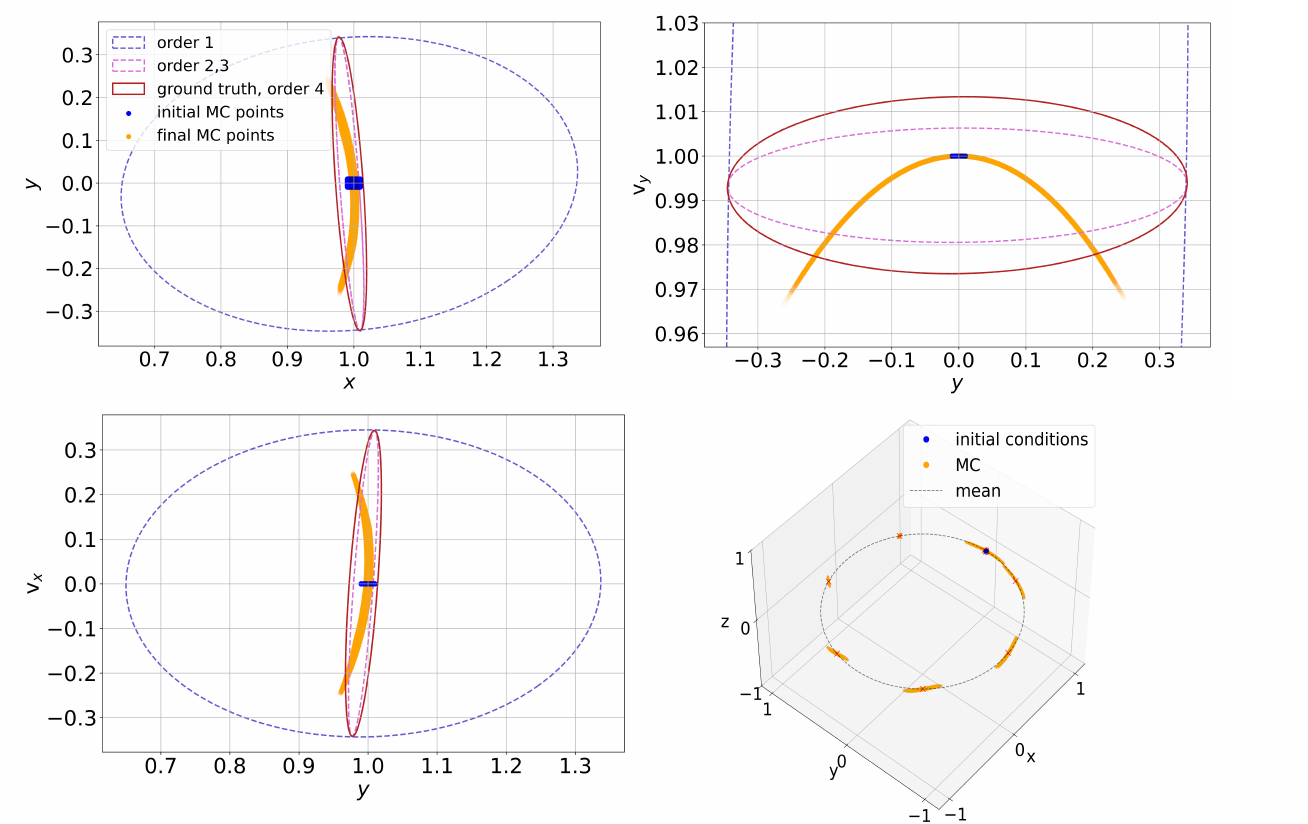}
  \caption{Monte Carlo and Taylor expansion approximation of the moments for the two body unperturbed cases.}
	\label{fig:two_body_plot}
\end{figure*}

\begin{figure*}[tbh!]
  \centering
\includegraphics[width=1\columnwidth]{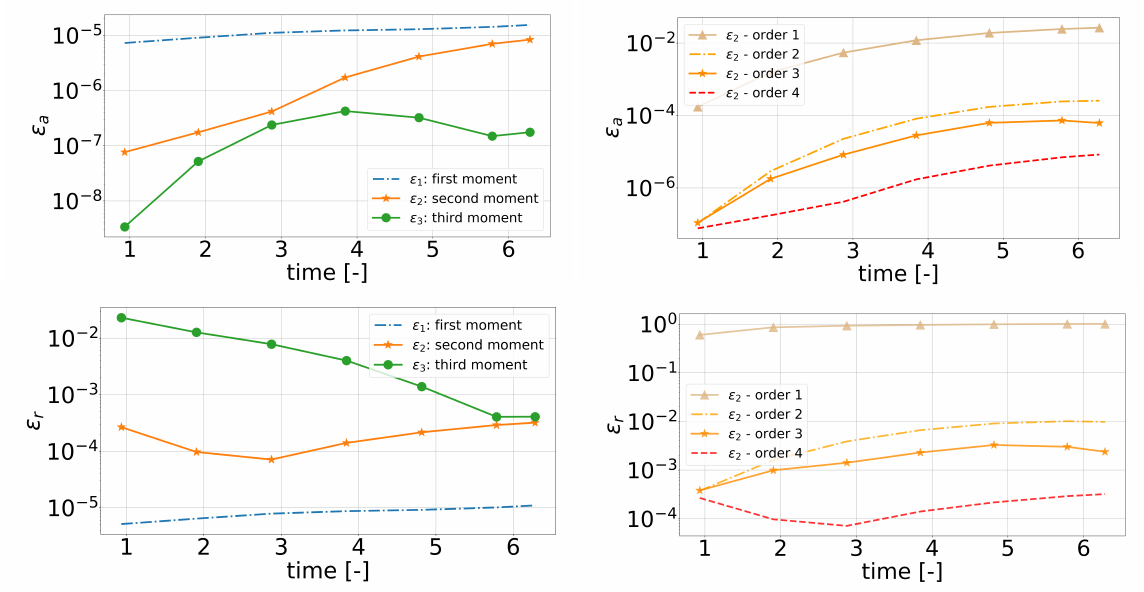}
  \caption{Errors as a function of time in the first three moments (left), and of the second moment for different orders (right).
  }
\label{fig:two_body_epsilon}
\end{figure*}

In Fig.~\ref{fig:two_body_plot}, we display 4 projections of the final reconstructed statistics, in terms of mean and covariance, using different Taylor order expansions of our proposed method, against the ones from the Monte Carlo method. In all figures, the order 1 estimation of the covariance is off, while orders 2 and 3 overlap and almost converge towards the ground truth distribution (the Monte Carlo one, in this case), and order 4 overlaps with the ground truth. Moreover, in the lower right plot, we also display (in orange) the three-dimensional Monte Carlo-sampled states as they are propagated along the orbit. As it can be seen, for all projections, from order 2 and above, the technique manages to very accurately capture both the mean and covariance matrix of the underlying probability density function. Furthermore, as we observe from the orange and blue samples, which represent the Monte Carlo trajectories before and after the integration time, respectively, the probability density function does not start from a Gaussian, nor converges towards it after the integration. To diagnose the non-normality of the propagated distribution, one can use moments above the second: for instance, the third moment provides information about the skewness of the distribution. In Fig.~\ref{fig:two_body_epsilon}, we display the accuracy as a function of integration times at which the proposed technique can reconstruct the first three moments at a 4th-order Taylor expansion (on the left), together with the accuracy at which the second statistical moment (i.e., the covariance) is reconstructed at different orders, ranging from the first to the 4th (on the right). The error in the reconstruction of the $k$-th statistical moment is computed via the absolute error:
\begin{equation}
    \varepsilon_{a,k}=||m_{k,est}-m_{k,t}||^2\text{,}
    \label{eq:epsilon_metric}
\end{equation}
as well as the relative error, defined as 
\begin{equation}
    \varepsilon_{r,k}=\varepsilon_{a,k}/||m_{k,t}||^2.
    \label{eq:relative_error}
\end{equation}
In both Eq. \eqref{eq:epsilon_metric} and \eqref{eq:relative_error}, we use $m_{k,t}$ to indicate the true $k$-th statistical moment tensor, which in this case is computed via Monte Carlo, and $m_{k,est}$ to indicate the approximated $k$-th statistical moment tensor, which in this case is computed by the method introduced in Sec.~\ref{sec:nonlinear_mapping_of_statistical_moments_of_perturbations}. For order one (i.e., $k=1$), the norm is a standard 2-norm of a vector, however, for higher order, this becomes the Frobenius norm (due to the fact that the moments become tensors). 
As observed from Fig.~\ref{fig:two_body_epsilon}, the method effectively captures the first three moments with relative errors below 1\% in the reconstruction of the moment tensors. The relative error is useful as it provides a direct interpretation in terms of percentage, allowing for an intuitive assessment of the reconstruction accuracy.
However, the utility of relative error diminishes when the denominator, i.e., the norm of the $k$-th order moment, approaches zero. In such cases, the relative error can become disproportionately large or even undefined, rendering it meaningless. 
To address this limitation, one can resort to the absolute error, which we also display in the top row of the figure. This remains meaningful in scenarios where the moment norm goes to zero.


\subsubsection{Perturbed Case}

As a more complex variant of the Keplerian problem discussed in the above section, we here tackle the J2-perturbed version of the two-body problem~\cite{sun2019nonlinear}. In this case, the satellites' equations of motion are:
\begin{align*}
    \dot{\pmb{r}}&=\pmb{v}\\
    \dot{v}_x&=-\frac{\mu}{r^3}x-\dfrac{3\mu J_2 R_e^2}{2r^5}\bigg( 1-\dfrac{5z^2}{r^2}\bigg)x\\
    \dot{v}_y&=-\frac{\mu}{r^3}y-\dfrac{3\mu J_2 R_e^2}{2r^5}\bigg( 1-\dfrac{5z^2}{r^2}\bigg)y\\
    \dot{v}_z&=-\frac{\mu}{r^3}z-\dfrac{3\mu J_2 R_e^2}{2r^5}\bigg( 3-\dfrac{5z^2}{r^2}\bigg)z    
\text{,}
\end{align*}
where $R_e=6378,137$ km is the Earth radius, $J_2$ is the Earth $J_2$ perturbation coefficient and $\mu$ is the Earth gravitational parameter. It is assumed that the initial $x_0$ and $y_0$ components of the state, as well as the $\mu$ and $J_2$ parameters, are uncertain and can be described by a uniform probability density function. In particular, the first two components of the state have mean values: $\overline{x}_0=6771.3560$ km and $\overline{y}_0=0$ km, with upper and lower bounds of $\pm100$ m; while the mean of the parameters are $\overline{J}_2=0.0010826$ and $\overline{\mu}=398600.4418$ km$^3$/s$^2$, with upper and lower bounds $\pm 5$\%. The rest of the initial state is described by a Dirac-delta function with mean: $z_0=0$, $v_{x,0}=0$, $v_{y,0}=7.523$ km/s, $v_{z,0}=1.525$ km/s. The state is integrated for a full orbital period (i.e., about 1.5 hours). 

\begin{figure*}[tbh!]
  \centering
  \includegraphics[width=1\columnwidth]{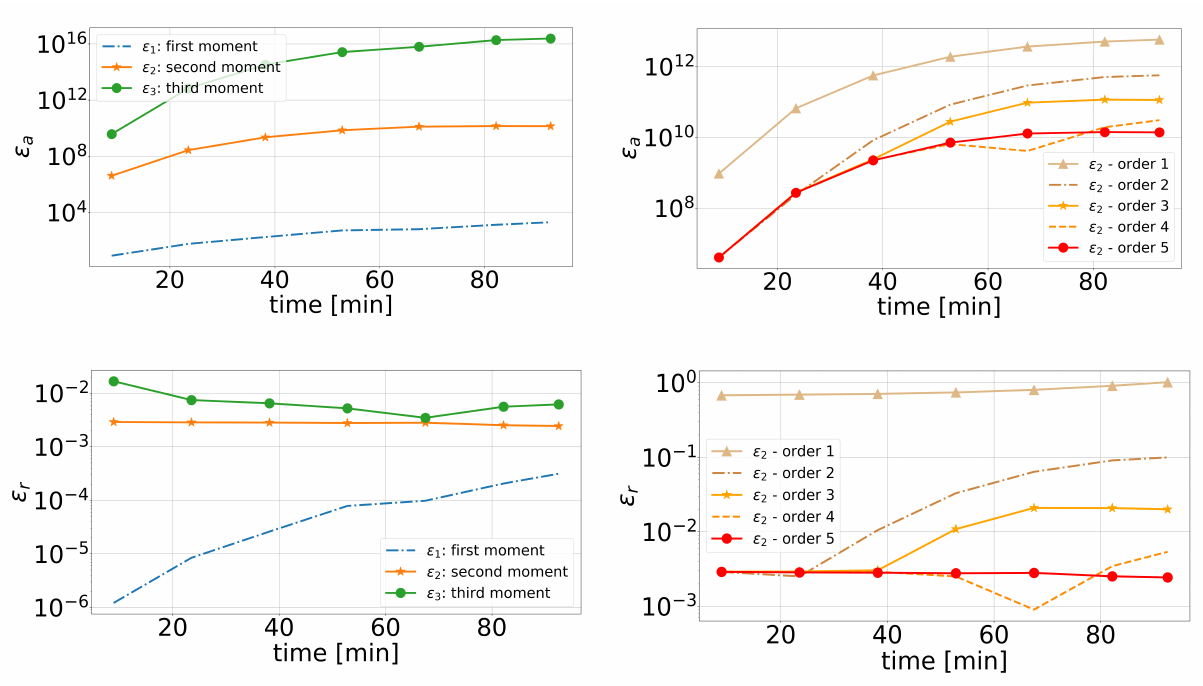}
  \caption{Errors in the reconstruction of the first three moments (left), and of the second moment for different orders, both as a function of time (right).}
\label{fig:two_body_perturbed}
\end{figure*}

As for the unperturbed case, also in this case the coupled equations of motion, cause the initial uncertainties in the $x,y$ component and in the parameters to eventually affect all six components of the state during the propagation. We use the Monte Carlo propagation with 10 million samples as ground truth, and we compare the reconstructed statistical moments with those found with the proposed approach, up to the Taylor expansion of order 5. As before, the time required to compute the Taylor expansion at order five of the state is orders of magnitude lower than the cost to perform the Monte Carlo. On the left of Fig.~\ref{fig:two_body_perturbed}, we display how the accuracy of the reconstruction in the first, second, and third moment behaves as a function of time, using the absolute and relative errors in the moment reconstruction discussed in Eq.~\eqref{eq:epsilon_metric}. The error is always maintained below 1\% of accuracy, for all the first three moments tensors. Furthermore, as seen from the right figure, order 1, which consists of just using the first-order state transition matrix, does not accurately reconstruct the second moment. Instead, increasing the order to two, three, and four, brings substantial improvements in the error of the reconstruction, while between order four and order five there is not a significant gap. 

\subsection{Circular Restricted Three-Body Problem}
\label{sec:cr3bp_experiments}
\subsubsection{Taylor Polynomial Convergence Radius of Different Families}
In this section, we investigate the use of the proposed uncertainty propagation method for mapping non-Gaussian uncertainties at the event manifold, in the context of the CR3BP. First, we will analyze the convergence radius for seven different periodic orbits, belonging to different families~\cite{howell2001families, restrepo2018database}. The equations of motion describing an object in motion (in non-dimensional units) around such a dynamical system can be described by~\cite{szebehely2012theory}:
\begin{align}
    \begin{split}
    \dot{\pmb{r}}&=\pmb{v}\\
    \dot{v}_x&=2v_y-\dfrac{\partial \overline{U}}{\partial x}\\
    \dot{v}_y&=-2v_x-\dfrac{\partial \overline{U}}{\partial y}\\
    \dot{v}_z&=-\dfrac{\partial \overline{U}}{\partial z}\text{,}
    \end{split}
    \label{eq:cr3bp_equations_of_motion}
\end{align}
where $\pmb{r}=[x,y,z]^T$ is the position vector, $\pmb{v}=[v_x,v_y,v_z]^T$ is the velocity vector, $\mu=m_2/(m_1+m_2)$ is the mass of the secondary body (e.g. Moon) divided by the mass of the primary (e.g. Earth) and secondary, with $m_1>m_2$, while $\overline{U}=-1/2(x^2+y^2)-(1-\mu)/r_1-\mu/r_2$ is the effective potential, $r_1^2=(x+\mu)^2+y^2+z^2$ and $r_2^2=(x+\mu-1)^2+y^2+z^2$.

In the context of the Earth-Moon system, we analyze two members of the Northern Halo L2 family (1-NH-L2 and 2-NH-L2), two members of the Southern Halo L2 family (1-SH-L2 and 2-SH-L2), one member of the distant retrograde orbit family (DRO), and one member of the Lyapunov vertical L1 family (Ly-L1). Then, we also analyze a member of the short-period L4 family (SP-L4), in the context of the Saturn-Enceladus system. 

In Tab.~\ref{table:cr3bp_orbits}, we report the values of the initial state, mass ratio parameters, period (all in non-dimensional units), and stability indices, for all the above-mentioned cases. We recall that for a periodic orbit in the CR3BP, its state transition matrix after exactly one period (known as monodromy matrix) has two eigenvalues equal to 1 (also known as trivial eigenvalues), and the remaining four are reciprocal in pairs (i.e., $\lambda_2=1/\lambda_1$ and $\lambda_4=1/\lambda_3$). 
\begin{table*}[h!]
\scriptsize
\centering
\caption{Initial conditions, period, mass ratio parameter, and stability for the CR3BP systems and periodic orbits used.}
\begin{tabular}{|c|c|c|c|c|c|c|c|c|c|c|}
\hline
 & $x_0$ & $y_0$ & $z_0$ & $v_{x,0}$ & $v_{y,0}$ & $v_{z,0}$ & T & $\mu$ & $|\textrm{Re}(\nu_1)|$ & $|\textrm{Re}(\nu_2)|$ \\ \hline
1-NH-L2 & 1.069 & 0.& 2.013$\times 10^{-1}$& 0.& -1.848$\times10^{-1}$& 0.& 2.174& 0.01215& 1.957& 0.180\\ \hline
2-NH-L2 & 1.159 & 0.& 1.255$\times10^{-1}$& 0.& -2.090$\times10^{-1}$& 0.&  3.266& 0.01215 & 355.101& 0.793\\ \hline
1-SH-L2 & 1.159 & 0.& -1.255$\times10^{-1}$& 0.& -2.090$\times10^{-1}$& 0.& 3.266 & 0.01215 & 355.101& 0.793\\ \hline
2-SH-L2 & 1.091& 0.& -2.014$\times10{-1}$& 0.& -2.092$\times10^{-1}$& 0.& 2.512& 0.01215 & 5.961& 1.655\\ \hline
DRO & 6.431$\times10^{-1}$& 0.& 0.& 0.& 8.177$\times10^{-1}$& 0.& 5.369& 0.01215 & 0.797& 1.662\\ \hline
Ly-L1 & 8.905$\times10^{-1}$& 0.& 0.& 0.& -3.611$\times10^{-1}$& -0.9773& 6.168& 0.01215 & 362.293& 17.183\\ \hline
SP-L4 & 5.0$\times10^{-1}$& 1.369& 0.& 7.705$\times10^{-1}$& -3.844$\times10^{-1}$& 0.& 6.283 & 1.90111$\times10^{-7}$& 2.& 2.\\ \hline
\end{tabular}
\label{table:cr3bp_orbits}
\end{table*}
The stability of the orbit is established by the non-trivial eigenvalues. In particular, two stability indexes are defined: $\nu_1=(\lambda_1+1/\lambda_1)$ and $\nu_2=(\lambda_3+1/\lambda_3)$, which can be used to assess the stability of the periodic orbit. For the orbit to be stable, the absolute value of the real part of both these indices must be below 2. If that is not the case, then the orbit can admit unstable and stable manifolds emanating from the periodic orbit~\cite{connor1984three}. The objective of these experiments is to represent a diverse set of periodic orbits in the CR3BP, with different stability properties, to study how the convergence radius of the Taylor expansion varies, depending on the dynamical environment surrounding these orbits. 

\begin{figure*}[tbh!]
  \centering
  \includegraphics[width=1\columnwidth]{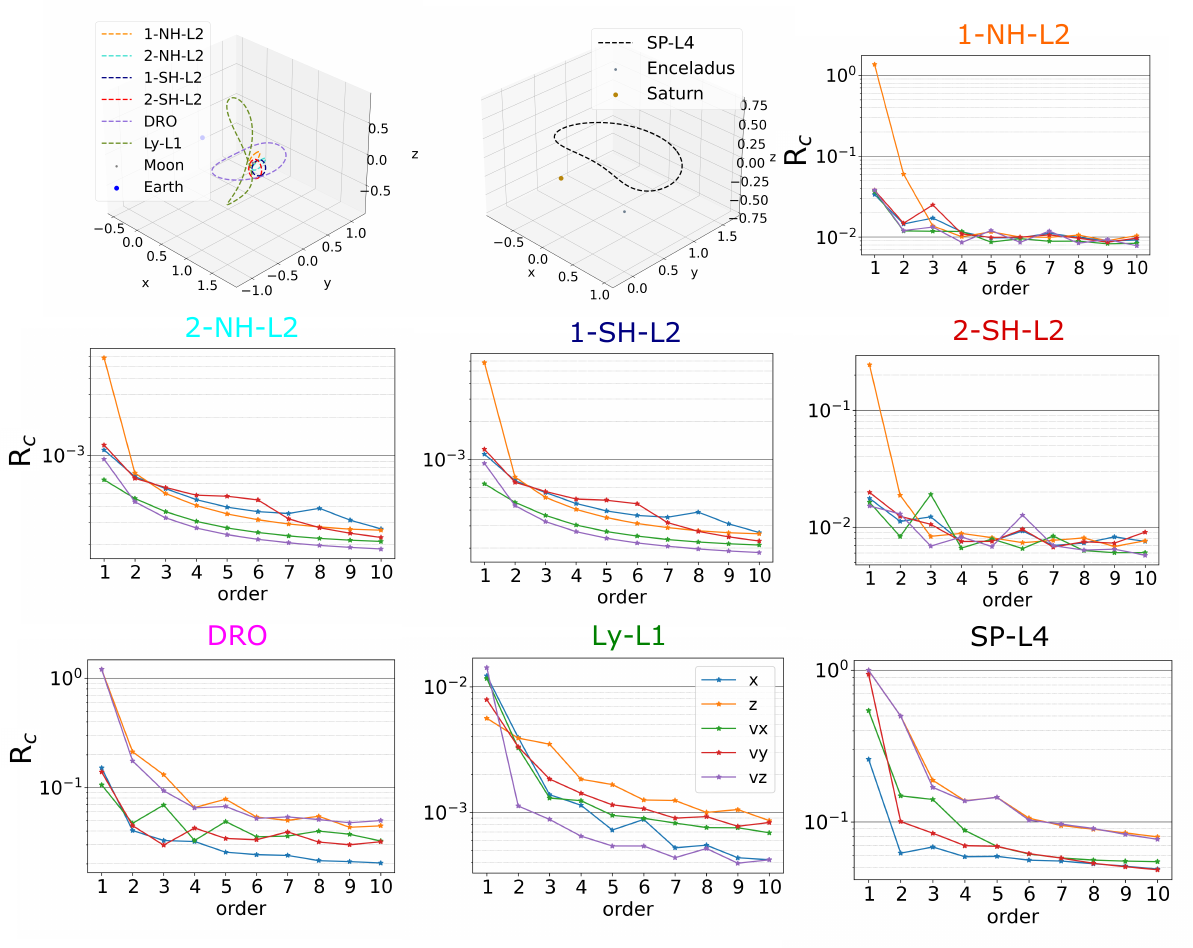}
  \caption{Three-dimensional CR3BP periodic orbits in the Earth-Moon (upper left) and Enceladus-Saturn (upper center) systems and Cauchy-Hadamard's radius of convergence estimation.}
\label{fig:cr3bp_ratio_tests}
\end{figure*}

\begin{figure*}[tbh!]
  \centering
 \includegraphics[width=1.\columnwidth]{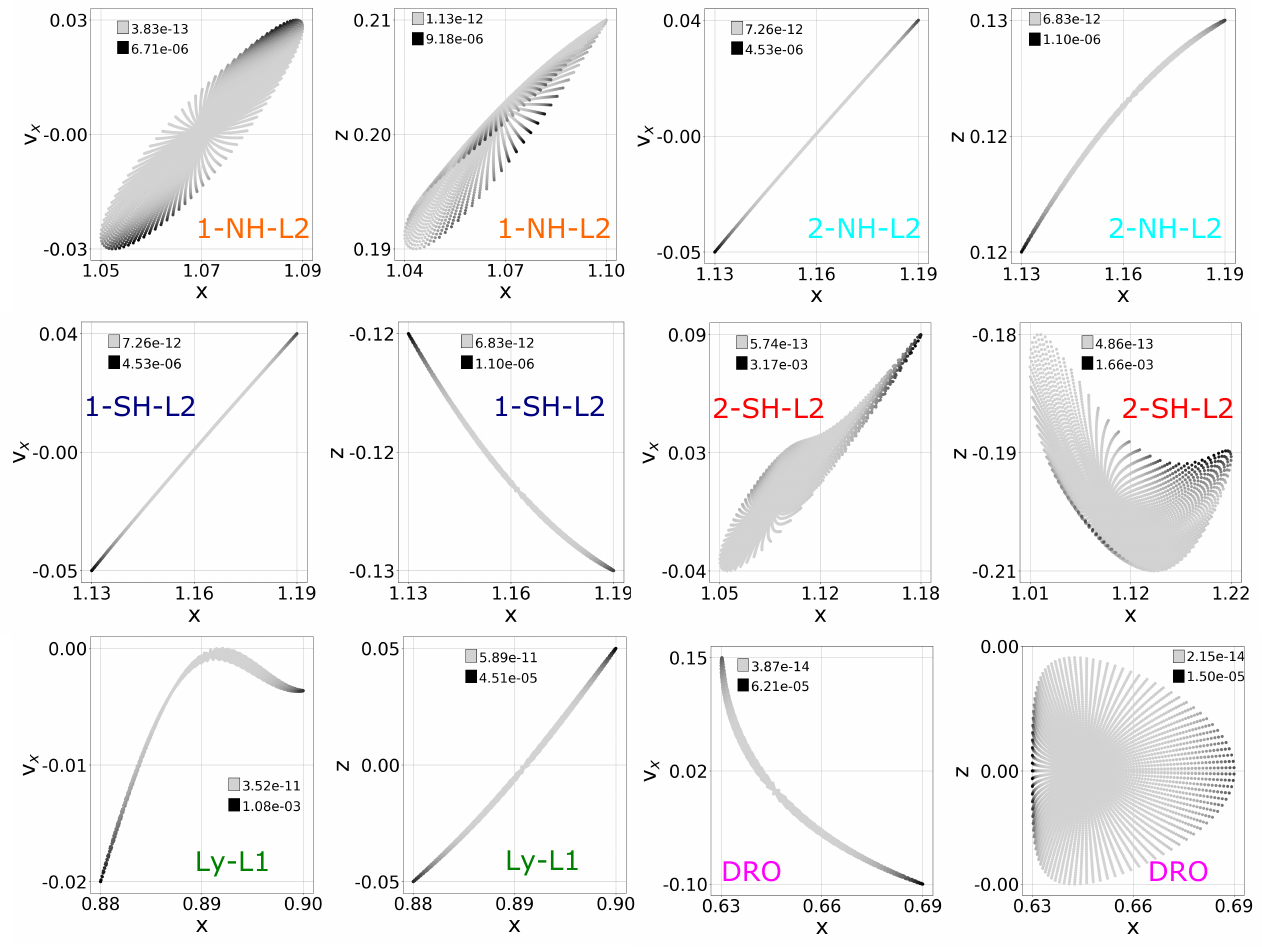}

  \caption{1\textsuperscript{st} and 3\textsuperscript{rd} column: numerically integrated Poincaré maps; 2\textsuperscript{nd} and 4\textsuperscript{th} column: error between the ground truth and the Taylor approximation.}
\label{fig:cr3bp_convergence_radius_earthmoon}
\end{figure*}

\begin{figure*}[tbh!]
  \centering
    \includegraphics[scale=0.17]{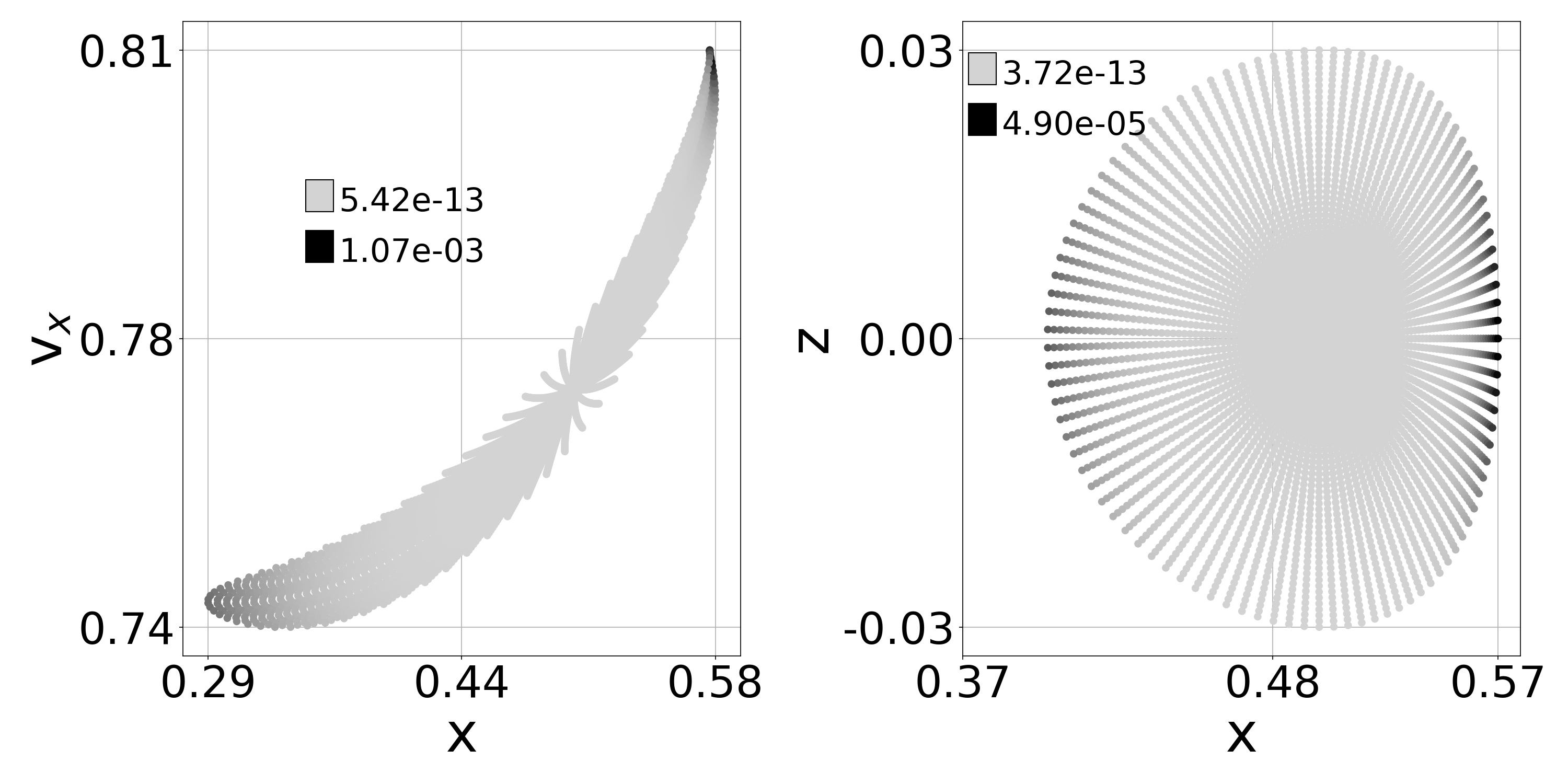}
  \caption{Ground truth and Taylor approximation of Poincaré map for a short-period L4 orbit in the Saturn-Enceladus system.}
\label{fig:cr3bp_convergence_radius_saturnenceladus}
\end{figure*}

In the two top left plots of Fig.~\ref{fig:cr3bp_ratio_tests}, we show the three-dimensional orbits around the Earth-Moon and Saturn-Enceladus system. Furthermore, in the remaining plot, we display the radius of convergence for each component of the state, as estimated by Cauchy-Hadamard's theorem, shown in Eq.~\eqref{eq:cauchy_hadamard}. 
To generate these, for each orbit, we construct the Taylor expansion after a period and compute the Cauchy-Hadamard radius of convergence for each component of the state. Then, the lowest radius across the different components estimated at the highest order can be used as a proxy of the ball radius in which the Taylor series converges. Hence, this guides the choice for the initial uncertainty size that can be acceptable. While this criterion is often used in practice, it is important to remember that it would only be valid if the Taylor series is expanded to infinite terms: truncating at lower orders only results in an approximation of the convergence radius. To verify that the criterion is working properly, in Fig.~\ref{fig:cr3bp_convergence_radius_earthmoon}, we also report two projections of the ground truth Poincaré map, propagated at different ball radii, together with the associated error w.r.t. the Taylor approximation, for all the different periodic orbits (each row corresponds to a periodic orbit). In the figure, we use the same color for ball radii of the same size. As we observe, the error of the Taylor approximation is maintained very low, when the perturbation is below the Cauchy-Hadamard's radius of convergence. We also perform the exact same analysis for the Saturn-Enceladus short-period L4 orbit, which we report in Fig.~\ref{fig:cr3bp_convergence_radius_saturnenceladus}. Also in this case, we confirm that for perturbations below the Cauchy-Hadamard's radius of convergence, the errors between the Taylor approximation and the ground truth orbit are maintained very low, thereby confirming that this criterion can be used as a good proxy for the region of validity of the Taylor expansion.
\subsubsection{Statistical Moments Mapping onto Poincaré Maps}

Having verified the accuracy of the Taylor reconstruction within the estimated Cauchy-Hadamard convergence radius, we perform the final step of our numerical experiments by mapping the initial statistical moments to the event manifold (in our case a Poincaré map) of interest, which in this case we define as $y_f-y_0=0$, starting from initial conditions corresponding to the periodic orbit of interest. For this numerical experiment, we use the 2-SH-L2 periodic orbit in Tab.~\ref{table:cr3bp_orbits}, assuming that the indicated mass ratio parameter has a uniform uncertainty with $\pm$1\% upper and lower bounds. We also assume that the $x_0, z_0, v_{x,0}, v_{y,0}$ components are uncertain and described by a uniform distribution with upper and lower bounds of $\pm10^{-4}$ (that is, roughly, $\pm38$ km in position, and $\pm0.1$ m/s in velocity). 

We first compute the relative error between the reconstructed and ground truth second order moment using Eq.~\eqref{eq:epsilon_metric}, obtaining for order 1 $\varepsilon_{r,2}=2.4017$, for order 2 $\varepsilon_{r,2}=0.0109$, for order 3 $\varepsilon_{r,2}=0.0106$, for order 4, and order 5, $\varepsilon_{r,2}=0.0105$. This confirms not only that the accuracy improves, as higher orders for the Taylor expansion are used, but also that for this case there is not an improvement between order 4 and order 5, which suggests that order 4 might have sufficed if the objective was the reconstruction of the second moment only.

Finally, to also have a quantitative estimation of how well the first three moments are reconstructed, we compute the relative error in the reconstruction of each moment with the Taylor expansion at order 5, using Eq.~\eqref{eq:epsilon_metric}. In this case, we obtain $\varepsilon_{r,1}=2.4665\times10^{-6}$, $\varepsilon_{r,2}=0.0105$, and $\varepsilon_{r,3}=0.0452$. This confirms that also in this case the first three moments can be reconstructed with a relative error below 0.001\% for the mean vector, around 1\% for the covariance ellipsoid, and below 5\% for the third moment.

\section{Conclusion}

In this Note, we present a novel approach for uncertainty propagation using high-order Taylor expansions of the flow and moment generating functions. While previous studies predominantly focus on Gaussian distributions, our method leverages the relationship between moment generating functions and distribution moments to extend high-order uncertainty propagation techniques to non-Gaussian scenarios, broadening their applicability across diverse problems and uncertainty types.

The proposed approach is general by design and suitable for arbitrary dynamical systems, as demonstrated through numerical simulations across various astrodynamics applications, including the unperturbed and perturbed two-body problem and the mapping of statistical moments within the framework of the circular restricted three-body problem. Our results reveal that the method effectively propagates non-Gaussian uncertainties with manageable computational costs and achieves low error rates compared to statistical moments generated via traditional Monte Carlo simulations.

Notably, high-order moment computation is performed once, using a symbolic manipulator independent of specific dynamics and probability density functions. This optimization reduces the computational burden of uncertainty propagation to the calculation of Taylor series coefficients around a nominal trajectory, which is efficiently executed through the integration of the system’s variational equations. The code supporting this work is available open-source\footnote{\url{https://gitlab.surrey.ac.uk/stag/up-with-mgf}, accessed July 2024}, to facilitate researchers to use and build upon these techniques.
\clearpage 
\bibliography{references}

\end{document}